\documentclass[prl,twocolumn]{revtex4}
\usepackage{graphicx, epsfig}
\usepackage{color}
\usepackage{mathrsfs}

\newcommand{\be}{\begin{equation}}
\newcommand{\ee}{\end{equation}}
\newcommand{\bea}{\begin{eqnarray}}
\newcommand{\eea}{\end{eqnarray}}
\newcommand{\ba}{\begin{eqnarray}}
\newcommand{\ea}{\end{eqnarray}}

\newcommand{\gapp}{\mathrel{\raise.3ex\hbox{$>$}\mkern-14mu
              \lower0.6ex\hbox{$\sim$}}}
\newcommand{\lapp}{\mathrel{\raise.3ex\hbox{$<$}\mkern-14mu
              \lower0.6ex\hbox{$\sim$}}}

\begin{document}
\title{Cosmology of Bifundamental Fields}

\author{Tanmay Vachaspati}
\affiliation{
Institute for Advanced Study, Princeton, NJ 08540\\ 
CERCA, Department of Physics, 
Case Western Reserve University, Cleveland, OH~~44106-7079
}

\begin{abstract}
\noindent
If a field theory contains gauged, non-Abelian, bi-fundamental 
fields {\it i.e.} fields that are charged under two separate
non-Abelian gauge groups, the transition from a deconfined phase 
to a hadronic phase may be frustrated. Similar frustration may
occur in non-Abelian gauge models containing matter only in
higher dimensional representations {\it e.g.} models with pure 
glue, or if ordinary quarks are confined by two flux tubes, as 
implied in the triangular configuration of baryons within QCD. 
In a cosmological setting, such models can lead to the formation 
of a web of confining electric flux tubes that can potentially 
have observational signatures. 
\end{abstract}

\maketitle

Many current theories of the fundamental interactions,
contain bi-fundamental fields 
(e.g. \cite{Franco:2008jc,Balasubramanian:2008tz})
that transform non-trivially under two separate non-Abelian
symmetry groups. The bi-fundamental nature arises in
string theory models because a string has two ends,
each of which is confined to a brane. The string
state bridging the two branes acts like a field that 
is charged under the symmetry groups corresponding to 
each of the two branes. Thus it is a bi-fundamental field. 
Here we will explore possible cosmological implications of 
such models. More specifically, we consider a field theory 
with symmetry group
\begin{equation}
[SU(N) \times SU(M)] \times {\rm (SM)}
\label{firstclass}
\end{equation}
where SM refers to the Standard Model groups and the
factors within square brackets will be referred 
to as the ``hidden sector''. Interesting cosmological 
considerations arise if we further assume 
\begin{itemize}
\item There is matter that transforms non-trivially
      under both $SU(N)$ {\it and} $SU(M)$.
\item $SU(N)$ and $SU(M)$ are both confining. 
\item Matter which is a singlet under either
      $SU(N)$ or $SU(M)$ (but not both) and in the 
      fundamental representation of the other group 
      does not exist or is very heavy compared to the
      confinement scale.
\end{itemize}
The class of models above can be trivially generalized to 
the case of more symmetry factors in the hidden sector. 
Our considerations also extend to models of the kind
\begin{equation}
[SU(N)] \times (SM)
\label{secondclass}
\end{equation}
provided
\begin{itemize}
\item There exist matter fields transforming in 
the adjoint (or higher dimensional) representations 
of $SU(N)$. 
\item The $SU(N)$ factor is confining.
\item Fields in the fundamental representation of the
$SU(N)$ factor are not present or are very heavy compared 
to the confinement scale.
\end{itemize}
To be concrete, we will mostly discuss models of the type in 
Eq.~(\ref{firstclass}). Though it should be noted that models
with only glue fall in the class in Eq.~(\ref{secondclass}).
Also, if each quark is connected to two confining flux tubes, 
as in the triangle model of baryons (e.g. \cite{'tHooft:2004he}), 
then QCD will fall into this class.

The simplest case to consider is with $N=M=2$ but we shall 
use $N=M=3$ for illustrative purposes. In the flux tube 
picture of confinement, each bi-fundamental is attached to 
two electric flux tubes, one to confine the $SU(N)$ flux 
and the other to confine the $SU(M)$ flux. Further, since 
there are no states in the fundamental representation, the 
flux tubes cannot break. 
The same holds true in the model in Eq.~(\ref{secondclass}), 
where each particle is confined by two (or more) flux tubes 
since the particle is assumed to be in a higher than minimal 
representation and the absence of fundamentals means that
the flux tubes cannot break.

\begin{figure}
  \includegraphics[width=2.0in,angle=0]{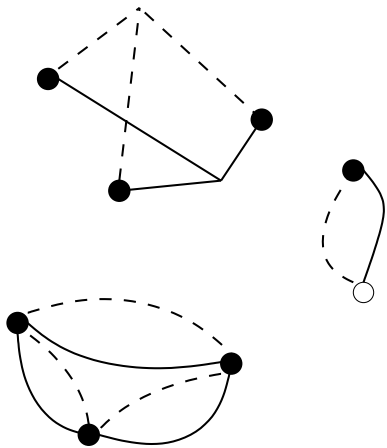}
\caption{Baryons consist of 3 or more bi-fundamentals (filled
black circle), each confined by two types of flux tubes.
The flux tubes of each type in a baryon may be in ``Y'' configuration
(upper drawing) or ``triangular'' configuration (lower drawing).
Mesons consist of particle-antiparticle.
}
\label{baryonmeson}
\end{figure}

The low enery excitations of the hidden sector of these theories 
include states similar to mesons and baryons in the standard
model. (We will continue to use standard model terminology,
{\it e.g.} hadrons, baryons, mesons etc., to refer to objects 
in the hidden sector when there is no confusion.) The hadrons 
are clusters of bi-fundamentals that are singlets of both 
$SU(N)$ and $SU(M)$. In the picture of particles and confining 
flux tubes they can be represented as in Fig.~\ref{baryonmeson},
where we show both the ``Y'' configuration and the 
``triangular'' configuration for the baryons. 
If $N=M=2$, the hadrons correspond to closed loops
of string beaded with bi-fundamental particles.

\begin{figure}
  \includegraphics[width=2.0in,angle=0]{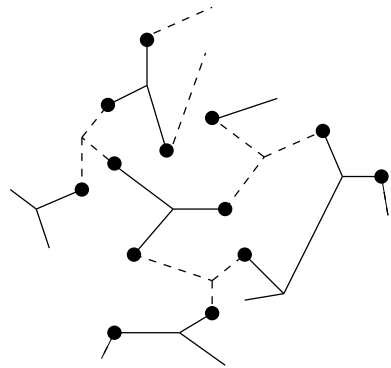}
\caption{The web of bi-fundamentals. Two types of flux tubes 
connect each particle, and the network percolates, forming an 
infinite web.
}
\label{net}
\end{figure}

Now let us consider the deconfined to confined phase transition. 
Our experience with the corresponding dual picture, where 
particles are replaced by magnetic monopoles and confining 
flux tubes by magnetic strings, suggests that such a transition 
is not possible. Instead, particles and strings form an 
infinite cosmic web as schematically represented in 
Fig.~\ref{net}. (In the case of $N=M=2$, the web is
replaced by a set of infinite strings with bi-fundamental
beads on them.) The system of confining flux tubes percolates, 
much as cosmic strings percolate at a phase transition, putting
almost all of the energy of the cosmic string network in
infinite strings \cite{Vachaspati:1984dz}. 
In other words, the transition from deconfined bi-fundamentals 
to a hadronic phase is ``frustrated''.
To make the transition to a purely hadronic phase, the flux 
tubes have to find very particular bi-fundamental particles 
to connect to, so that the entire web can break up into hadrons. 
There are many more ways to connect the bi-fundamentals so that 
the structure is that of a web. Even though the lowest energy
state contains only baryons and mesons, the lowest energy
state is also one of very low entropy and is hard to arrive at.
The larger $N$ and $M$ are, the greater is the frustration,
and the denser is the network.

A direct way to see that a web, and not a gas of baryons and 
mesons, is the likely outcome, is to note in Fig.~\ref{net}
that vertices form where 3 flux tubes come together. The 
bi-fundamental particles are simply junctions between two 
different types 
of flux tubes and may be ignored for the purpose of the 
web structure. Then the flux tubes form a network that 
is just like a network of $Z_3$ strings.  Simulations of 
$Z_3$ string network formation show that $> 90\%$ 
of the string is in one infinite network \cite{Aryal:1986cp}.
It may also be possible to study the formation of a web 
in the dual picture where magnetic monopoles carrying several 
different non-Abelian charges get confined by strings due to 
the breaking of large non-Abelian symmetries. Such a possibility 
is discussed in the context of grand unified models in 
\cite{Ng:2008mp}.

A second view of the frustration is in terms of the competition
between energy and entropy. In a system containing strings without
junctions, the low temperature {\it equilibrium} state consists of
a distribution of closed loops, which are the analogs of hadrons.
At high temperature it is known to be favorable to put all
the string energy into infinite strings. The temperature at 
which long strings become favored is called the Hagedorn temperature 
\cite{Hagedorn:1965st} and occurs at the temperature where the 
entropy contribution to the free energy becomes more important than 
the Boltzmann suppression \cite{Frautschi:1971ij,Carlitz:1972uf,
Mitchell:1987hr,Mitchell:1987th}. The Hagedorn transition has 
also been found for strings with junctions \cite{Rivers:2008je}.
Note that the Hagedorn picture is based on a system in thermal 
equilibrium. It is a separate question as to whether interactions 
occur sufficiently rapidly that they can maintain equilibrium. At 
a phase transition, the interactions become too slow to maintain 
equilibrium and it becomes possible for topological defects and, 
in our case, a cosmic web to survive as a remnant.
This is similar to the freeze-out of heavy particles that is so 
crucial to the existence of cosmic dark matter.

The Hagedorn picture has been directly confirmed in studies of $U(1)$ 
cosmic string formation \cite{Antunes:1997pm} 
(also see \cite{Sakellariadou:1987kc}) where a $U(1)$
system in equilibrium at high temperature is gradually cooled
down resulting in the production of infinite structures.
Similarly in the case of bi-fundamentals, as the universe cools,
it may become energetically favored for the system to break
up into baryons and mesons but, before this can happen, the
web falls out of equilibrium and the transition to the hadronic
phase is frustrated. The formation of a web will occur even
if the number density of bi-fundamental particles is very low,
as low as one per horizon volume. 

It is an interesting question if a web can form in the total 
absence of bi-fundamentals. Then the model only contains
gluons and, as noted above, falls into the class in
Eq.~(\ref{secondclass}). The system at high temperature 
will contain gauge particles of the non-Abelian groups, 
$SU(N)$ for example. As the system is cooled below the
deconfining temperature, all the gauge particle must get
confined by two flux tubes each since they are in the adjoint
representation. Once more we expect an infinite $Z_N$ web,
not a gas of glueballs, to form.

In the QCD literature, there is discussion of whether baryons
are better modelled by quarks connected by confining strings
that are in a Y configuration or a triangular ($\Delta$) 
configuration. Fig.~\ref{baryonmeson} can be used to illustrate 
these configurations if we remove one type of confining flux
tube e.g. the dashed lines. A crucial difference between the 
Y and $\Delta$ configurations is that in the former case each 
quark is connected to one flux tube, whereas in the latter case 
each quark is connected to two flux tubes. Just as for the
bi-fundamentals, if during the QCD phase transition, each 
quark is connected to two flux tubes, we expect hadronization 
to be frustrated because the strings will percolate and
form infinite structures. Only a fraction of quarks will 
then end up as hadrons. The rest will form beads on a 
network of QCD cosmic strings that contain color
electric fields. The structure of the QCD string network
will then be like the $N=M=2$ case of Eq.~(\ref{firstclass}).
However, it is not clear if the $\Delta$ model of the baryons 
should be taken literally and applied to a quark plasma. In the 
dual picture, where the quarks are replaced by magnetic monopoles 
and the confining flux tubes by Nielsen-Olesen strings carrying 
color magnetic flux, each monopole is connected to only one 
string.

Going back to bi-fundamental models,
despite the conclusive evidence that an infinite network
forms in the dual (magnetic) model, there is no proof that 
such networks must exist in the electric sector. Hence it
is useful to review the ingredients that lead to the
existence of the infinite (magnetic) network. In the 
magnetic sector, the vacuum expectation value of an 
order parameter spontaneously breaks a symmetry. 
The order parameter lies on a ``vacuum manifold'' and 
the choice of point on the vacuum manifold is uncorrelated 
beyond a certain distance. 
The finite correlation length of the order parameter also 
follows generally from causality arguments \cite{Kibble:1976sj}
and implies that any topological structures ({\it e.g.} 
strings) that are formed are oriented in random fashion 
beyond a certain distance. Hence, a string is approximately 
described by a random walk \cite{Vachaspati:1984dz}.
Now random walks in three spatial dimensions are known
not to close and so the strings typically formed during
a phase transition are infinite in extent. Infinite 
strings correspond to infinite networks when the strings 
can have junctions. In the electric sector too, we are
discussing a phase transition from the deconfined to
confined phase, and the phase transition must be described
by an order parameter. It seems reasonable to assume that
the order parameter lies on some non-trivial manifold.
Causality implies that the order parameter takes on 
uncorrelated values beyond a certain distance and
hence any strings that are formed are also randomly 
oriented. Then, just as in the magnetic sector, we
expect infinite networks to be present in the electric
sector too. The situation appears similar to that in
string theory where it is possible to have cosmic
networks in both magnetic and electric sectors
\cite{Klebanov:2000hb,Copeland:2003bj}.

Once a web freezes out, it can only relax due to the usual 
dynamical factors --
tension in the strings, cosmic expansion, interactions of 
strings, and annihilation of particle-antiparticle. 
Flux tubes belonging to the same symmetry group intercommute 
on intersection, while flux tubes of different type 
will pass through each other. If the tension in one kind 
of string is larger than the other, the dynamics will cause
the larger tension strings to shrink while stretching out
the lower tension strings. Then the network will evolve
toward a web of just one kind of string 
\cite{McGraw:1997nx,Copeland:2005cy}. In Fig.~\ref{net}, 
if the strings shown by dashed lines have larger tension,
they shrink and bring together 3 bi-fundamentals
to form a singlet of (say) $SU(M)$ which is now a vertex 
for a $Z_3$ network of light strings.

The energy density in the network as compared to that in other 
standard model particles is an important quantity. Work on 
cosmic string networks \cite{Vachaspati:1986cc,McGraw:1997nx,
Copeland:2005cy,Hindmarsh:2006qn,Urrestilla:2007yw}
suggests that if there are efficient
energy loss mechanisms, it is reasonable to expect that the 
energy density of the web will scale with cosmic expansion 
{\it i.e.} the fraction of energy density in the web remains 
constant in a given cosmology. (The constant may change in 
transitions such as from radiation to matter to dark energy 
domination.) Possible energy loss channels for the web include 
production of mesons, baryons, closed string configurations 
(``glueballs''), and gravitational waves. If there are interactions
between the hidden sector and the standard model particles,
the web could also decay into photons and other light standard
model particles. 

We now summarize the properties of the cosmic web following
Ref.~\cite{Vachaspati:1986cc}. Cosmological consequences of
a frozen network of strings have also been investigated in 
Refs.~\cite{Spergel:1996ai,Bucher:1998mh}.

The canonical scenario is that the bi-fundamentals do not carry 
any charges other than the confined $SU(N)$ and $SU(M)$ charges. 
Then there is no efficient energy loss mechanism for the web and 
the cosmological evolution is that of a fluid with equation of 
state parameter, $w$, that takes into account the energy-momentum 
for both the strings and the bi-fundamental particles. The pressure, 
$P_{\rm web}$, and energy density of the web, $\rho_{\rm web}$,
are related by
\begin{equation}
P_{\rm web} = \frac{\rho_{\rm web}}{3} 
         [ \beta + (1-\beta) (2 \langle v_s^2 \rangle -1 ) ]
\equiv \gamma \rho_{\rm web}
\label{eqofstate}
\end{equation}
where $\beta \in (0,1)$ is the fraction of the web energy in 
(relativistic) bi-fundamentals and $\langle v_s^2 \rangle$ is 
the average squared velocity of the strings. Since
$0 < \langle v_s^2 \rangle < 1$, the equation of state
parameter is constrained by $ -1/3 < \gamma < 1/3$.
Recent studies of string networks indicate 
$2 \langle v_s^2 \rangle -1 < 0$ \cite{Copeland:2006if}.

\begin{figure}
\centerline{\scalebox{0.75}{\input{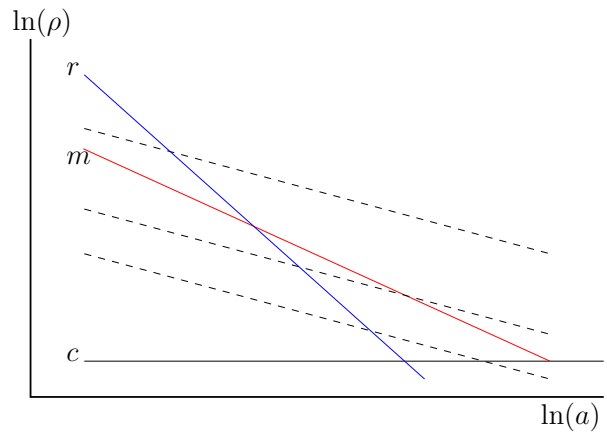}}}
\caption{The plot shows the schematic evolution of cosmic 
energy densities for the radiation (r), matter (m), and
cosmological constant (c). The web energy density is
shown by the dashed lines, each line assuming a different
initial density. If the initial energy density in the
web is larger than that of the matter component, the
universe enters a web dominated epoch and never a 
matter dominated epoch. With smaller initial energy density
(middle dashed line), the web can dominate at recent epochs, 
prior to cosmological constant domination. With yet smaller
initial energy density, the web remains sub-dominant
throughout.
}
\label{evolution}
\end{figure}

The cosmological evolution follows from solving the 
Friedmann-Robertson-Walker equations with four components
to the energy density: radiation, matter, web, and
cosmological constant (or dark energy), with equation
of state parameters 
$w = 1/3, 0, \gamma , \ {\rm and}\ -1$ 
respectively. The energy densities
in the components decay with scale factor as
$a^{-4}, a^{-3}, a^{-3(1+\gamma )} , \ {\rm and}\ a^0$. 
The radiation density decays the fastest.
Assuming $\gamma < 0$, which will
happen if the web is dominated by strings that are
not too relativistic, the matter density decays
faster than the web density, while the cosmological constant
energy density does not decay. If we start in a radiation 
dominated universe, depending on the initial energy density 
of the web, evolution can lead to early web domination in
conflict with observation, or to a recent epoch of web
domination that may be cosmologically acceptable. These
possibilities are shown schematically in Fig.~\ref{evolution}.


To determine the initial energy density of the web, we note
that confinement will set in when the separation of bi-fundamental 
particles becomes of order of the inverse confinement scale:
$n^{-1/3} \sim \Lambda^{-1}$ where $n$ is the number
density of bi-fundamentals and $\Lambda$ the confinement
scale. So the energy density in the web at the confinement
scale is $\rho_w (t_{\rm c}) \sim m \Lambda^3 + \mu \Lambda^2$
where $m$ is the mass of a bi-fundamental and the string
tension $\mu \sim \Lambda^2$. If the bi-fundamental mass
is less than the confinement scale ($m < \Lambda$), the
energy density in the network is dominated by strings and
$\rho_w (t_{\rm c}) \sim \Lambda^4$. The cosmic temperature
at formation is $T_c \sim \Lambda$ and the critical 
cosmic energy density $\sim \Lambda^4$. Hence the web initially 
contains an $O(1)$ fraction of the cosmic energy density.
If $m > \Lambda$, the bi-fundamentals become non-relativistic 
at some high temperature and start red-shifting like 
pressureless matter. When the annihilation cross-section
drops below the Hubble expansion rate, they freeze-out of 
equilibrium. The exact freeze-out density depends on the 
exact interactions but, in addition, freeze-out may be more 
complicated than for standard dark matter because once the 
separation between bi-fundamentals grows to $\Lambda^{-1}$, 
the bi-fundamentals get confined by flux tubes. 

Another scenario considered in \cite{Vachaspati:1986cc} 
corresponds to one in which the bi-fundamentals also carry
$U(1)$ gauge charges, effectively making them
``tri-fundamentals''. Then the web has an efficient energy
loss mechanism due to radiation of gauge quanta and
the web does not come to dominate the matter density and
instead scales at a fixed ratio to the matter energy 
density.

The web scaling hypothesis implies that the
average distance between strings in the web grows linearly 
with cosmic time, $d = \chi t$, where $\chi$ is a constant. 
If we assume a thermal density of bi-fundamentals such that
the separation is the confinement scale $\Lambda^{-1}$, as
in the $m < \Lambda$ case discussed above,
we have $\chi = 1/(\Lambda t_c)$ where $t_c$ is the epoch
at which the confinement transition is supposed to occur.
The cosmic temperature is also given by $\Lambda$: 
$T(t_c) \approx \Lambda$. Therefore $t_c = T_P/\Lambda^2$ where
$T_P$ is the Planck temperature. Hence $\chi = \Lambda/T_P$.

Assuming that the web energy is dominated by non-relativistic
strings,
the web energy density is $\rho_{\rm web} \sim \mu/d^2$.
The cosmic critical energy density is ${\bar \rho}= 3/(8\pi Gt^2)$
and so the fraction of cosmic energy density in the web 
is constant at the value
\begin{equation}
\Omega_{\rm web} \sim \frac{8\pi G\mu}{3 \chi^2} \sim 1 
\end{equation}
This result is sensitive to the assumed value of $\chi$ but 
it is interesting to see that a cosmologically significant 
amount of energy density may reside in a bi-fundamental web. 

The scenario where a tangled web of strings plays the role of the 
observed dark energy has been considered in 
\cite{Spergel:1996ai,Bucher:1998mh}.
Current observations indicate an equation of state parameter 
$w \lesssim -0.8$ which is well outside the range for the web 
equation-of-state parameter. This implies that the web cannot 
explain all of the observed cosmological acceleration but 
it may still be a component of the cosmic energy density. As 
shown in Fig.~\ref{evolution}, the web could be the dominant 
energy component after last scattering and before cosmological 
constant domination {\it i.e.} at cosmological redshifts larger 
than a few. In this scenario, large-scale structure growth 
would slow down during the string-dominated epoch and this
constrains the redshift at which the web can start dominating, 
$z_{\rm sd} \lesssim 1$, so that there is sufficient time
available to obtain non-linear structures from density
perturbations $\delta \rho / \rho \sim 10^{-3}$ at last
scattering ($z \approx 1000$). 

The web does not affect the dynamics of standard model particles 
through particle interactions since it lies in the hidden sector
of particle physics. The exception is that the web still has 
gravitational interactions with the cosmological medium. In 
particular, in Ref.~\cite{Starkman:2000gu} 
it was pointed out that the interaction of a tangled network of 
strings with black holes at the centers of galaxies would displace 
the black holes and may possibly be used to constrain the energy
density in the web. 
 
Another possibility is that a dense web may be a candidate for the
dark matter. It should be noted that the strings in the web 
themselves are expected to be relativistic but the coarse grained, 
root-mean-squared velocity of the web may still be non-relativistic, 
in which case the web could be a candidate for cold dark matter 
(``confined cold dark matter''). The gravitational clustering 
properties of the web deserve further investigation. 

Finally, for completeness, we consider the case when a 
heavy fundamental field is present. Then a confining string
can break due to the nucleation of a pair of fundamental particles.
We may expect the whole network to break up into hadrons on a time 
scale set by the nucleation rate of fundamental particles on strings. 
If this time scale is shorter than the Hubble time, the network
will break up; if it is larger, we may expect the web to survive.
The breaking process is essentially that of Schwinger pair 
production and the breaking rate per unit time per unit
length of string is \cite{Vilenkin:1982hm,Monin:2008}
\begin{equation}
\frac{d\Gamma}{dl} = C \frac{\mu}{2\pi} e^{-\pi m_f^2 /\mu }
\end{equation}
where $m_f$ is the mass of the fundamental and $C \sim 1$. 
With Hubble expansion, the length of string in a Hubble
volume grows and eventually the breaking rate becomes 
faster than the Hubble rate. The network decays at cosmic 
time $t_*$ given by,
\begin{equation}
H(t_*) = l(t_*) \frac{d\Gamma}{dl} 
  = t_* \frac{\Lambda^2}{2\pi} e^{-\pi m_f^2 /\Lambda^2 }
\end{equation}
where $l(t_*) \sim t_*$ is the total length of string in a
Hubble volume at $t_*$ and $\mu = \Lambda^2$.
Further, with $H(t_*) \approx 1/t_*$, we get
\begin{equation}
t_*  \sim \Lambda^{-1} ~ e^{+\pi m_f^2 /(2 \Lambda^2 ) }
\end{equation}
Depending on parameters, interesting cosmological scenarios are 
possible. For example, there may be a period of string domination 
followed by the rapid break-up of the web, leading to
the production of other hidden sector particles. It may also
happen that the web breaks up relatively early, producing
hadrons that are out of equilibrium and whose energy density 
then redshifts like matter. In this scenario, there is the
danger that the bi-fundamental hadrons will start to dominate
the universe too early, as in the old cosmological magnetic 
monopole over-abundance problem \cite{Preskill:1979zi}.

To conclude, we have shown that bi-fundamental fields that
occur in current high energy physics models may lead to a
network of (electric) cosmic strings. 
Depending on the string tension and the density of the network, 
the web may be probed and constrained by observations. 
For example, in some cases, the network may dominate the 
cosmological energy density during the early universe or
affect galactic dynamics. If the network energy density 
scales due to the emission of quanta, these could be 
non-gravitational signatures of the web. Such consequences 
can lead to constraints on model building. Further work 
on the evolution of string networks within the context of 
specific models is needed to further explore the cosmological 
consequences of field theories containing bi-fundamental fields.

I am grateful to Vijay Balasubramanian, Joe Polchinski,
Nathaniel Seiberg, Al Shapere, Tomar Volansky, Brian Wecht, 
and Edward Witten for useful comments and discussion. I also
acknowledge support by the U.S. Department of Energy at Case 
Western Reserve University, grant number DE-FG02-90ER40542 at
the Institute for Advanced Study, and a grant from 
The Foundational Questions Institute (fqxi.org).

\end{document}